\documentstyle[12pt]{article}
\def\beq{\begin{equation}}
\def\eeq{\end{equation}}

\begin{document}

\begin{titlepage}
\begin{center}
{\Large \bf Theoretical Physics Institute \\
University of Minnesota \\}  \end{center}
\vspace{0.15in}
\begin{flushright}
TPI-MINN-93/20-T \\
May 1993
\end{flushright}
\vspace{0.2in}
\begin{center}
{\Large \bf Non-perturbative production of multi-boson states
and quantum bubbles \\ }
\vspace{0.15in}
{\bf A.S. Gorsky \\}
Institute of Theoretical and Experimental Physics,
Moscow, 117259 \\
and \\
{\bf M.B. Voloshin  \\ }
Theoretical Physics Institute, University of Minnesota,
Minneapolis, MN 55455 \\
and \\
Institute of Theoretical and Experimental Physics,
Moscow, 117259 \\
\vspace{0.4in}
{\bf   Abstract  \\ }
\end{center}
The amplitude of production of $n$ on-mass-shell scalar bosons
by a highly virtual field $\phi$ is considered in a $\lambda \phi^4$ theory
with weak coupling $\lambda$ and spontaneously broken symmetry. The
amplitude of this process is known to have an $n!$ growth when the produced
bosons are exactly at rest.  Here it is shown that for $n \gg 1/\lambda$ the
process goes through `quantum bubbles', i.e. quantized droplets of a
different vacuum phase, which are non-perturbative resonant states of the
field $\phi$.  The bubbles provide a form factor for the production
amplitude, which rapidly decreases above the threshold.  As a result the
probability of the process may be heavily suppressed and may decrease with
energy $E$ as $\exp (-const \cdot E^a)$, where the power $a$ depends on the
number of space dimensions. Also discussed are the quantized states of
bubbles and the amplitudes of their formation and decay.

\end{titlepage}

\section{Introduction}

The problem of calculating amplitudes of production of a large number $n$ of
weakly interacting bosons has received a close attention in connection with
the observation$^{\cite{ringwald,espinosa,mattis}}$ that the
instanton-induced amplitudes in the standard electroweak model, when
calculated to lowest orders of the perturbation theory (in the instanton
background), display a rapid growth with energy, associated with the growing
multiplicity of gauge and Higgs bosons in the final state. Subsequently
Cornwall$^{\cite{cornwall}}$ and Goldberg$^{\cite{goldberg}}$ have pointed
out that a similar growth takes place for the amplitudes of processes, in
which many bosons are produced by few initial particles in a simpler setting
of a $\lambda \phi^4$ theory of a scalar field $\phi$. The reason for this
growth is that in an amplitude of a process involving large number $n$ of
weakly interacting bosons the smallness of the coupling constant is
compensated by a large number of perturbation theory graphs, which typically
grows as $n!$. This growth is a manifestation of the well known
factorial divergence of the coefficients of the perturbation
theory$^{\cite{zinn}}$, and thus for $n > O(1/\lambda)$, at which $n$ the
compensation takes place, the perturbation theory becomes unreliable.

In this paper we consider the amplitudes of production of $n$ slow
on-mass-shell bosons by a virtual field $\phi$, `$1 \to n$ process', in a
theory of one real scalar field with spontaneously broken symmetry with
respect to the reflection $\phi \to -\phi$. The Lagrangian of this theory
in Minkowski space-time has the well known form

\beq
{\cal L}={1 \over 2} (\partial_\mu \, \phi)^2 - {\lambda \over 4} (\phi^2-
v^2)^2~~,
\label{lagr}
\eeq
where $\lambda$ is a small coupling constant and
$v$ is the vacuum expectation value of the field. The mass $m$
of the bosons propagating in either of the vacua at $+v$ or $-v$ is
$m=\sqrt{2 \lambda}\, v$. A number of exact results has been obtained
recently related to the amplitudes of the $1 \to n$ process at the
threshold, i.e. when the final $n$ particles are produced at rest. The sum
of the tree graphs for these amplitudes was originally found
explicitly$^{\cite{akp}}$ by using recursion relations$^{\cite{v}}$ :

\beq
\langle n |\, \phi(0) \, | 0 \rangle = n!\, (-2 v)^{1-n}
\label{a0}
\eeq
and then reproduced within a functional technique, suggested by
Brown$^{\cite{brown}}$, an extension of which technique will be heavily
used throughout this paper. Within this extension the problem of calculating
the threshold production amplitudes for the theory with the Lagrangian
(\ref{lagr}) reduces to a Euclidean space-time calculation of the
quantum average of the field $\phi$ with the kink-type boundary conditions
in (Euclidean) time: $\phi \to -v$ at $t \to -\infty$ and $\phi \to +v$ at
$t \to +\infty$. The tree-level expression (\ref{a0}) for the threshold
production amplitudes is determined by the coefficients of expansion in
powers of $e^{mt}$ at $t \to -\infty$ of the classical kink profile

\beq
\phi_0({\bf x},t) = v \tanh (mt/2)~~,
\label{phi0}
\eeq
which provides the saddle point for the classical action $S[\phi]$.

The first correction to the result (\ref{a0}) which amounts to summing
one loop graphs$^{\cite{v2}}$ has been found and
reduces$^{\cite{smith}}$ to multiplying the lowest order expression
(\ref{a0}) by the factor $(1 + c(d)\,n (n-1)\, \lambda)$, where $c(d)$ is a
coefficient depending on the number of space dimensions $d$,

\beq
\langle n |\, \phi(0) \, | 0 \rangle_{tree+1loop} = n! \,(-2 v)^{1-n}\,
(1 + c(d)\,n (n-1)\, \lambda)~~.
\label{r1}
\eeq
The $n^2 \lambda$ behavior of the relative magnitude of the correction
suggests that at large $n$ the saddle point configuration in calculation
of the $n$-th coefficient of the expansion at $t \to -\infty$ of the mean
field, i.e.  of the quantity $\phi(x) e^{-S}$, is driven far away from the
saddle point configuration (\ref{phi0}) of the action alone.
Once the correct saddle point configuration is found, the quantum
fluctuations around it should produce only subleading in $n$ corrections.(A
similar situation in a (1+1)-dimensional toy model was discussed in Ref.
\cite{v4}.)

Here we find that in the limit of large $n$ the correct saddle point
configuration in calculation of the threshold production amplitudes for
large $n$ is determined by dynamics of the surface of the inter-phase
boundary (`domain wall') separating the phases with $\phi = -v$ at large
negative $t$ and with $\phi=+v$ at large positive $t$. A WKB
treatment of this dynamics relates it to the semiclassical properties of
the spherical bubbles filled with the phase $\phi = +v$ in the vacuum with
$\phi=-v$. At the classical level such configurations were studied some time
ago$^{\cite{kobzarev}}$. In particular it was found that these
configurations are reasonably long-lived: a large bubble undergoes several
pulsations of its radius before decaying into outgoing waves.

In what follows it will be shown that in a theory in $(d+1)$ dimensions
(thus $d$ being the number of {\it spatial} dimensions) at $n \gg 1/\lambda$
the WKB result for the amplitude $\langle n |\, \phi(0) \, | 0 \rangle$ with
all particles having exactly zero momenta can be interpreted as given by a
two-stage process. The field operator $\phi(0)$ produces a bubble and then
the bubble couples to $n$ particles. The resulting amplitude is large:

\beq
\langle n |\, \phi(0) \, | 0 \rangle \sim n! (-2v)^{1-n} \exp( b(d)\, n^{d
\over {d-1}})
\label{asymp}
\eeq
with positive coefficient $b(d)$ depending on $d$. This growth is in
agreement with the general result$^{\cite{v3}}$ that in this theory the
threshold amplitudes should grow not slower than as $n!$. However exactly at
the threshold the phase space of the final $n$ particles is vanishing and to
estimate the total probability of the process one needs to know the behavior
of the amplitude above the threshold. The presence of the bubble in the
intermediate state implies existence of a form factor, which cuts off the
amplitude above the threshold when any of the momenta of the final particles
is larger than the inverse of the radius of the bubble, $r^{-1}$.  An
estimate of the total probability of the process $1 \to n$ with the
two-stage picture can be done by evaluating the probability of creation of a
bubble.  The probability of creation of a bubble with energy $E$ evaluated
by means of the Landau-WKB technique$^{\cite{lwkb}}$ is found in this paper
to be given by

\beq
|\,\langle B(E) | \phi(0) | 0 \rangle\,|^2 \sim \exp (-2\,b(d)\, (E/m)^{d
\over {d-1}})
\label{bublp}
\eeq
with precisely the same coefficient $b(d)$ as in eq.(\ref{asymp}), and $
B(E)$ stands for the state of the bubble with energy $E$.

Equation (\ref{bublp}) shows that due to the form factor provided by the
bubble not only the growth of the probability with energy is eliminated but
in fact the total probability associated with multi-boson states rapidly
falls with energy in the non-perturbative asymptotic regime. Therefore we
conclude that it is extremely plausible that in this theory the
non-perturbative contribution to the processes with production of many soft
final particles by few initial ones does not become large at high energy in
spite of the indications to the contrary in lowest orders of the loop
expansion. Whether this behavior is universal and is valid for other
theories, in particular for the electroweak theory, is yet to be studied.

The suppression of the non-perturbative processes of the type $few \to many$
at high energy does not contradict to a possible growth with energy of the
probability of the processes $many \to many$, as is discussed in the
concluding section.

The rest of this paper is organized as follows. In Section 2 a brief review
is given of the standard perturbative calculation of the amplitudes of the
$1 \to n$ processes within the equivalence to the problem of calculating the
quantum average of the field with the kink-type boundary conditions in the
Euclidean space. In Section 3 the problem of finding the proper saddle point
configuration for calculation of the amplitudes is formulated within the
so-called thin wall approximation. The search for this configuration leads
to considering dynamics of bubbles in the Minkowski space-time. This
dynamics is discussed within the Bohr - Sommerfield quantization in Section
4. The properties of bubbles are quantitatively related to the $n$ particle
production amplitudes in Section 5 and in Section 6 the probability of
creation of a bubble by a highly virtual field is estimated by the
Landau-WKB method. Section 7 contains a discussion of the results and of the
validity of the approximations made in present calculations.

\section{Perturbative calculation}

In this section we recapitulate the perturbative calculation of the
threshold production amplitudes within the equivalence$^{\cite{v3}}$ of this
problem for the theory described by the Lagrangian (\ref{lagr}) to the
Euclidean-space calculation of the quantum mean field with boundary
conditions corresponding to a domain wall separating vacua $+v$ and $-v$.
Fixing for definiteness that the amplitudes are calculated in the `left'
vacuum, i.e. at $\langle \phi \rangle = -v$ the equivalent problem
can be formulated as follows.

One first calculates in the Euclidean space-time the quantum mean field

\beq
\Phi(t) = {{\int \left ( \int \phi( {\bf x}, t)\, d^dx \right ) \,
e^{-S[\phi]}\,{\cal D} \phi} \over { V \, \int \, e^{-S[\phi]}\, {\cal D}
\phi}}~~,
\label{mf}
\eeq
where $d$ is the number of {\it
spatial} dimensions in the problem, $V$ is the $d$-dimensional spatial
normalization volume, and the boundary condition in the path integral  at
$t \to -\infty$ is specified by the requirement that the asymptotic behavior
of $\Phi(t)$ there is given by

\beq
\Phi(t) \to -v + z e^{mt} + O(e^{2 m t})
\label{bc}
\eeq
with $z$ being a constant and $v$ and $m$ are the renormalized v.e.v. and
the boson mass. The production amplitudes are then given$^{\cite{brown,v3}}$
by the coefficients of the expansion of $\Phi(t)$ in powers of $e^{mt}$ at
large negative $t$, i.e. if one writes the expansion as

\beq
\Phi(t)= \sum_{n=0}^{\infty} c_n \, e^{nmt}~~,
\label{series}
\eeq
then

\beq
\langle n | \phi(0) | 0 \rangle = n!\, c_n /z^n~.
\label{ac}
\eeq
The coefficient $z=c_1$ in the leading asymptotic behavior (\ref{bc}) is
thus a normalization of a one particle state: without dividing by the
factor $z^n$ in eq.(\ref{ac}) one would obtain the amplitudes with the
particle states normalized as $\langle 1 | \phi(0) |0 \rangle = z$. As it is
written the equation (\ref{ac}) gives the amplitudes with the standard
normalization of the particle states. It is also assumed here that the
normalization volume $V$ is set to unity.

The application of the described equivalence is straightforward at the tree
level, which corresponds to substituting for the field $\phi$ the
uniform in space (but not time) solution (\ref{phi0})
of the classical Euler-Lagrange equations for the Lagrangian (\ref{lagr}).
Expanding this solution in powers of $e^{mt}$ reproduces through
eq.(\ref{ac}) the tree-level result (\ref{a0}) for the production
amplitudes.

The solution (\ref{phi0}) is the familiar kink profile of the inter-phase
boundary (`domain wall') separating the vacua $-v$ at $t \to -\infty$ and
$+v$ at $t \to +\infty$. A particular choice of the normalization factor $z$
fixes the position of the domain wall in time, i.e. fixes the
translational zero mode of the field, which corresponds to an overall shift
in the time direction. Therefore one can either choose a
particular value of $z$ or alternatively fix the overall position of the
field in time. Throughout this paper we
choose to fix the position in time by requiring that the integration in the
path integral in eq.(\ref{mf}) runs over field configurations such that
$\phi({\bf x},0)$ is zero at the boundary of the normalization spatial
bounding box. At the classical level this fixes the center of the domain
wall at $t=0$ and thus also sets $z=2v$.
The quantum corrections to the mean field (\ref{mf}) in
general renormalize the factor $z$.

To account for the quantum effects in the mean field (\ref{mf}) one writes
the full field $\phi$ as a sum of classical and quantum parts

\beq
\phi({\bf x}, t)=\phi_0({\bf x},t) + \phi_q ({\bf x}, t)
\label{cq}
\eeq
and evaluates the mean value of the quantum part of the field by
perturbation theory in the background field $\phi_0$. At the one-loop level
this calculation$^{\cite{v2,smith}}$ leads to the result in eq.(\ref{r1}).

\section{Semiclassical configurations for quantum effects at large $n$}

As is discussed in the Introduction the growth with $n$ of the loop
corrections within the quantization around the saddle point
configuration (\ref{phi0}) of the action indicates that at large $n$ the
$n$-th coefficient of the expansion of $\Phi(t)$ in powers of $e^{mt}$ at $t
\to -\infty$ is contributed by a semiclassical configuration which strongly
differs from that in eq.(\ref{phi0}). To find this appropriate `distorted'
configuration we first consider the one shown in Fig.1, where the
domain wall (the surface corresponding to $\phi({\bf x},t)=0$) assumes a
non-flat shape slowly varying with ${\bf x}$.  The surface can be described
by its ${\bf x}$-dependent deviation from $t=0$, i.e. by the solution
$t=-h({\bf x})$ of the equation $\phi({\bf x},t)=0$. Assume now that one
fixes the shape of the boundary corresponding to a particular function
$h({\bf x})$ and minimizes the action with respect to all other variables of
the field by solving the Euler-Lagrange equations with the surface of zeros
being fixed.  Then for large negative $t$ at the point ${\bf x}_0$
corresponding to the maximum of $h({\bf x})$ the $n$-th harmonics of the
field is given in the leading exponential approximation by

\beq
2v (-1)^{n-1} \exp \left (n m(t+h({\bf x}_0) + O(nm\delta) \right ) \sim
2v (-1)^{n-1} \exp (n m h({\bf x}_0)) e^{nmt}~.
\label{rh}
\eeq
This behavior can be understood by considering that at the point ${\bf x}_0$
the evolution of the field in time from $\phi=0$ towards $\phi = -v$
proceeds over the time $| t| -h ({\bf x}_0) + \delta $, where $\delta$
reflects the uncertainty related to the curvature of the surface of the
inter-phase boundary. This uncertainty is of a subleading importance
in situations where $h({\bf x}_0)$ is large, which as will be seen is the
case for $n \gg 1/\lambda$.

Equation (\ref{rh}) tells that the coefficient $c_n$ gets multiplied by
the factor $\exp (n m h({\bf x}_0))$. Thus in the leading WKB approximation
this coefficient can be evaluated as

\beq
c_n \sim \max \left [ \exp \left ( nm h({\bf x}_0) - S[h] + S_0 \right )
\right ]~,
\label{wkbc}
\eeq
where $S[h]$ is the action for the field
configuration described by the shape $h({\bf x})$ of the inter-phase
boundary and $S_0 = S[h=0]$ is the action of the unperturbed classical
solution (\ref{phi0}), and all the pre-exponential factors are omitted.

The appearance of the action $S_0$ with the plus sign in eq.(\ref{wkbc}) is
due to the fact that in the expression (\ref{mf}) for the mean field the
path integrals in both numerator and the denominator are calculated with the
kink-type boundary conditions. Thus the factor $\exp (S_0)$ appears from the
saddle point value of the denominator.

If $h({\bf x})$ varies at scale larger than the thickness of the wall,
$m^{-1}$ the action $S[h]$ can be calculated in the thin-wall
approximation$^{\cite{vko}}$ as the surface tension of the wall $\mu$ times
its area $A$:

\beq
S[h] = \mu A[h]~,
\label{twa}
\eeq
where

\beq
\mu = \int_{-\infty}^{+\infty} \left [ {1 \over 2} \left ( {d \over {dt}}
\phi_0 (t) \right ) ^2 + {\lambda \over 4} (\phi_0^2-
v^2)^2~~ \right ] \, dt = {2 \over 3}\, \sqrt {2 \lambda} \, v^3~ = {{m^3}
\over {3 \lambda}}.
\label{mu}
\eeq

Finding the maximum of the expression (\ref{wkbc}) with the action
(\ref{twa}) is equivalent to solving a surface tension problem for a
$d$-dimensional film in $(d+1)$-dimensions. The edges of the film are fixed
at the boundary of the bounding box: $h(boundary)=0$, and at the point ${\bf
x}_0$ the force equal to $nm$ is applied downwards. The maximal deviation
$h_0=h({\bf x}_0)$ of the film will be largest if the force is applied to the
center of the film, therefore we set ${\bf x}_0=0$. (In fact for $d > 2$ the
equilibrium shape of the film does not depend on ${\bf x}_0$ if this point
is sufficiently far from the edges. For $d \le 2$ there is an infrared
behavior in this problem, so that the equilibrium deviation explicitly
depends on the size of the bounding box. Also it is explicitly assumed
throughout this paper that $d > 1$. A one-dimensional `film' lacks
intrinsic curvature, which makes most of the formulas in this paper
singular in the formal limit $d \to 1$, i.e. that of a (1+1) - dimensional
field thory, for which the present analysys is thus not directly
applicable.) Assuming that the bounding box is spherically symmetrical in
the spatial $d$ dimensions with a large radius $R$, one concludes that the
shape, which the film takes under the force applied at the center, is also
spherically symmetrical and can be characterized by the radius $r(t)$ of its
slice at $t=const$ if the slice is positioned at an instant $t$ such that
$-h(0) < t < 0$.  The boundary conditions for $r(t)$ being $r(-h_0)=0$ and
$r(0)=R$.

In terms of $r(t)$ the quantity $S[h]-n\,m\, h_0$ entering the expression
(\ref{wkbc}) can be written as

\beq
s[r]=\int_{-h_0}^0  l_d \, \mu \, r^{d-1}\, \sqrt {1+ \dot{r}^2} \, dt - E
h_0 ~~,
\label{rl}
\eeq
where the factor $nm$ is identified with the total energy $E$,
$\dot{r}=dr/dt$, and $l_d$ is the $(d-1)$ dimensional volume of unit sphere
$S_{d-1}$:  $l_d=2 \, \pi^{d/2}/\Gamma (d/2)$. The integral in eq.(\ref{rl})
is nothing else than the Euclidean action for a spherical bubble in the thin
wall approximation$^{\cite{vko}}$ in the theory with degenerate vacua. Since
$E$ is the conserved value of the Hamiltonian for the classical trajectory
$r(t)$, the functional $s[r]$ is identified as the truncated action

\beq
s[r]= \int p_E\, dr ~,
\label{ta}
\eeq
where $p_E$ is the Euclidean momentum conjugate of $r$:

\beq
p_E={{ l_d \, \mu \,r^{d-1} \dot{r}} \over \sqrt{1 + \dot{r}^2}}~.
\label{mom}
\eeq
On an Euclidean-space classical trajectory the value $E$ of the Hamiltonian
is related to $p_E$ and $r$ as

\beq
E^2+p_E^2=( l_d \, \mu \, r^{d-1})^2 ~.
\label{cl}
\eeq

The solution of the latter equation for $p_E$ in terms of $E$ and $r$ reads
as

\beq
p_E = \sqrt{( l_d \, \mu \, r^{d-1})^2 ~ - E^2} =  l_d \, \mu \,
\sqrt{r^{2d-2}-r_0^{2d-2}}~~,
\label{pe}
\eeq
where

\beq
r_0=(E/( l_d \, \mu))^{1 \over {d-1}}~.
\label{r0}
\eeq
This solution immediately reveals an important point: there is no real
solution for $p_E$ and thus for $r(t)$ at $r < r_0$. In the equivalent
surface tension problem the origin of this behavior is obvious: there is a
minimal radius equal to $r_0$ of a slice of the surface with surface tension
$\mu$ that can support the force $E$. In terms of quantum mechanics with the
action (\ref{ta}) the point $r_0$ corresponds to the classical turning
point, and for $r < r_0$ the evolution of the system proceeds in the
Minkowski time. We are thus compelled to consider the evolution of the
radius of the bubble along the complex time trajectory shown in Fig.2, on
which the part of the trajectory with $r < r_0$ evolves along imaginary
Euclidean, i.e. real Minkowski, time. In the Minkowski space this part of
the trajectory describes the bubble expanding from size $r=0$ to the
classical turning point $r=r_0$.  Therefore before proceeding with further
evaluating the coefficients $c_n$ by eq.(\ref{wkbc}) it is
appropriate to discuss few properties of the bubbles in the Minkowski
space-time.

\section{Quantizing the bubbles}

The Hamiltonian dynamics of the bubble in the Minkowski space-time in the
thin wall approximation is described by the simple substitution $t \to it$
in the previous Euclidean space formulas. In particular the analog of the
equation (\ref{cl}) for the Hamiltonian is

\beq
H^2-p_M^2=( l_d \, \mu \, r^{d-1})^2
\label{clm}
\eeq
with the Minkowski momentum $p_M$. The classical Minkowski-space trajectory
with energy $E$ corresponds to oscillations of the bubble between the
turning point $r=r_0$ (eq.(\ref{r0})) and $r=0$.
Naturally it could be that instead of oscillating the bubble would quickly
dissipate into outgoing waves. However a numerical study of the classical
evolution of the field of the bubble-type configuration$^{\cite{kobzarev}}$
(not constrained by the thin wall approximation) has revealed that the
bubbles undergo at least several oscillations before they emit a larger
portion of their energy in outgoing waves. This implies that the lifetime of
a bubble is at least longer than the period of oscillation $T \sim r_0$.
Therefore we start with discussing the bubbles as if they were stable and
later take into account their slow decay.

The part of the trajectory near
zero radius, $r < m^{-1}$, cannot be described within the thin wall
approximation since the thickness of the wall is of order $m^{-1}$. However,
at large energy, corresponding to the turning radius $r_0 \gg m^{-1}$, most
of the evolution of the bubble proceeds within the applicability of the thin
wall approximation, and this is the part from which most of the action
comes.  According to equation (\ref{r0}) the condition $r_0 \gg m^{-1}$
implies that

\beq
n=E/m \gg m^{3-d}/\lambda~,
\label{cond}
\eeq
where the right hand side is the inverse of the dimensionless coupling in
the theory and thus is assumed to be much bigger than 1. In particular for
$d=3$ (the normal (3+1) dimensional theory) the condition
(\ref{cond}) translates into $n \gg 1/\lambda$.  Thus the applicability of
the present calculations lies within an essentially non-perturbative domain.

The oscillatory motion of the bubbles can be quantized and the discrete
energy levels found by applying the Bohr - Sommerfield quantization rule:

\beq
I(E) \equiv \oint p_M \, dr - 2\pi \, \nu(E)= 2\pi \, N~,
\label{bs}
\eeq
where the integral runs over one full period of oscillation and contains the
momentum $p_M$ determined by eq.(\ref{clm}) in the thin wall approximation.
The quantity $\nu(E)$ is a correction to the thin wall limit, which arises
from the contribution to the action of the motion at short distances $r \sim
m^{-1}$, where the latter limit is not applicable. Since at such
distances $p_M \sim E$, by order of magnitude $\nu(E)$ can be estimated as
$\nu(E) \sim E/m$. The integral in eq.(\ref{bs}) is of the order of
$E\,r_0$, and is thus much larger than $\nu(E)$ once the condition $r_0 \gg
m^{-1}$ is satisfied. In terms of the turning radius $r_0$ the quantization
relation (\ref{bs}) reads as

\beq
k_d \mu r_0^d = 2 \pi \, (N+\nu(E))~,
\label{bsr}
\eeq
with $k_d$ being a numerical coefficient,

\beq
k_d= l_d \, {{\sqrt{\pi}\, \Gamma[1/(2d-2)]} \over {2 (d-1) \, \Gamma[3/2 +
1/(2d-2)]}}~.
\label{kd}
\eeq
At a large energy, corresponding to the condition (\ref{cond}), the
correction term with $\nu(E)$ can be neglected and one finds the expression
for an energy level $E_N$ in terms of the number $N$ of the level:

\beq
E_N=\mu^{1 \over d}\, \left ( {{2 \pi \,N} \over {k_d}} \right )^{{d-1}
\over d}~.
\label{bse}
\eeq

By the relation (\ref{bsr}) the condition $r_0 \gg m^{-1}$ requires that

\beq
N \gg m^{3-d}/\lambda~,
\label{cn}
\eeq
where the right hand side by itself is assumed to be a large number. The
energy $E_N$ of the level then satisfies the condition (\ref{cond}). It can
be readily noticed that for such $N$ the spacing between consecutive levels
is small in comparison with the mass $m$ of the bosons: $\Delta E_N \sim
r_0^{-1} \ll m$. This, perhaps, in addition to the reflectionless property
of the wall explains the relative stability of large bubbles. Indeed,
emission of individual quanta would require transitions between states with
a large difference of their quantum numbers $N$: $\Delta N \simeq m\, r_0
\gg 1$. The overlap integral for the wave functions of the levels with
large difference of their numbers is exponentially small in $\Delta N$.
Therefore the probability of emission of the bosons by large bubbles should
be strongly suppressed by the parameter $m r_0$. However there perhaps can
be a larger probability of simultaneous emission of many bosons at short
distances, i.e. when the bubble contracts to $r < m^{-1}$ during the
oscillations. Admittedly  at present
there is little understanding of the decay of the bubbles
besides the study$^{\cite{kobzarev}}$ by means of classical field
equations.

\section{Quantized bubbles and multi-boson processes}

Returning now to evaluation of the coefficients $c_n$ by eq.(\ref{wkbc}), we
notice that the exponent there receives a real contribution from the action
on the Euclidean part of the trajectory, i.e. when $r$ changes between
$r=r_0$ and $r=R$, and an imaginary part on the Minkowski part of the
trajectory, i.e. between $r=0$ and $r=r_0$. Let us first evaluate the
Euclidean part, where the truncated action (eq.(\ref{ta})) is given by

\beq
s[r]=\int_{r_0}^R p_E \, dr =  \int_{r_0}^R \sqrt{( l_d \, \mu\, r^{d-1})^2
- E^2} \, dr~.
\label{seu}
\eeq
The latter integral diverges in the
limit $R \to \infty$.  However the quantity of interest is the difference
$S_0 - s[r]$, which enters the exponent in eq.(\ref{wkbc}). Since the action
$S_0$ of the unperturbed configuration formally corresponds to the truncated
action $s[r]$ at zero energy $E=0$ and has the same leading divergence, one
can first calculate the derivative of $s[r]$ with respect to $E$ and then
find the difference of the action as

\beq
S_0-s[r]= \int_0^E \left ( - { {d s[r]} \over {dE}} \right ) \, dE~.
\label{dife}
\eeq
For the derivative with respect to energy one finds from eq.(\ref{seu})

\beq
- { {d s[r]} \over {dE}} = {E \over { l_d \, \mu}} \int_{r_0}^R {{dr} \over
\sqrt{r^{2d-2}-r_0^{2d-2}}}~.
\label{da}
\eeq

Before discussing this integral for large energy, it can be noted that
quantitatively this expression can be used to estimate the quantum effects
in the coefficients $c_n$ also at low energies where formally $r_0$
is less than thickness of the wall. Then the integral should be cut off at a
lower limit $r=r_1 \sim m^{-1}$, which does not depend on $E$ if $E/m \ll
m^{3-d} /\lambda$. Then one finds

\beq
c_n \simeq c_n^{(0)} \exp \left ( {{E^2} \over {2 l_d \, \mu}} \int_{r_1}^R
{{dr} \over {r^{d-1}}} \right )~,
\label{appr}
\eeq
where $c_n^{(0)}$ is the tree-level expression for the coefficient $c_n$.
When the exponent is expanded in powers of its argument this gives the
$(1+ const \cdot n^2 \lambda)$ behavior of the first quantum correction to
the production amplitudes, which one finds by a calculation of
one-loop graphs$^{\cite{v2,smith}}$ (eq.(\ref{r1})). Moreover for $d \le 2$
the integral in equations (\ref{da}) and (\ref{appr}) still diverges in the
limit $R \to \infty$. Therefore the dependence on the infrared cut off $R$
dominates the quantum effects and makes them insensitive to the region of
small $r$. In this case in the leading at large $R$ approximation the
exponent in eq.(\ref{wkbc}) can be found for any $E$ with the obvious result

\beq
c_n=c_n^{(0)} \exp \left ( {{E^2 \, R^{2-d}} \over {2 (2-d) l_d \, \mu}}
\right ) = c_n^{(0)} \exp \left ( {{n^2 \, R^{2-d} \, \lambda} \over {6
\,(2-d) \, l_d \, m}} \right )~.
\label{ir}
\eeq
It is also clear that the Minkowski part of
the trajectory $r(t)$ contains no infrared dependence and thus does not
contribute to this result in the leading in $R$ approximation.
It can be readily verified by the technique of Refs.\cite{v2,smith} that the
infrared behavior of the one-loop corrections to the coefficients $c_n$
exactly reproduces the expansion in the latter equation of the exponent  up
the first power of its argument.

Therefore considering the case where the equation (\ref{da}) is applicable
down to $r_0$ with the condition that $r_0 \gg m^{-1}$ is sensible only when
the integral in that equation is finite in the limit $R \to \infty$, i.e.
when $d > 2$, which includes the most interesting case of $d=3$.
Setting $R = \infty$ in eq.(\ref{da}) one finds for $d > 2$

\beq
- { {d s[r]} \over {dE}} = f_d \, r_0(E) = f_d \, \left ( {E \over {l_d \mu}
} \right )^{1 \over {d-1}}
\label{sea}
\eeq
with the dimensionless factor
$f_d$ given by

\beq
f_d={{\sqrt{\pi} \, \Gamma[1/2- 1/(2d-2)]} \over {2 (d-1) \, \Gamma[1-
1/(2d-2)]}}~.
\label{fd}
\eeq

Therefore one finds the real part of the difference $S_0 - s[r]$ associated
with the Euclidean part of the trajectory $r(t)$ to be given by

\beq
(S_0-s[r])_E= f_d \,{{d-1} \over d}\, E \, \left ( {E \over {l_d \mu} }
\right )^{1 \over {d-1}}~.
\label{sef}
\eeq

Let us now evaluate the contribution to the coefficients $c_n$ of the
Minkowski part of the evolution of the bubble. The simplest trajectory,
which links $r=0$ with $r=r_0$ consists of one half of the period which
contributes to the coefficient $c_n$ the factor

\beq
F_0=\exp ( i \, I(E)/2 ) ~,
\label{halft}
\eeq
where the action integral $I(E)$ over a full period is defined in
eq.(\ref{bs}) However there are also trajectories connecting $r=0$ with
$r=r_0$, which differ from this one by an integer number $k$ of full periods
of oscillation.  Each of these trajectories provides a saddle point.
Therefore one has to take the sum of contributions of these saddle points.
The sum has the form

\beq
F(E)= \sum_{k=0}^\infty \exp [i\, (k+1/2) \, I(E)] =
{{\exp(i I(E)/2)} \over {1-\exp(i I(E))}}
\label{sumt}
\eeq
It is clear that this expression sums to infinity when the Bohr-Sommerfield
relation (eq.(\ref{bs})) is satisfied, i.e. $I(E)=2\pi\,N$. In other words
the factor $F(E)$ develops poles at the values of energy coinciding with the
positions of the bubble levels $E_N$ given by eq.(\ref{bse}).  It can be
reminded that thus far the decay of the bubble levels is completely ignored.
Therefore we come to the conclusion that in this approximation the amplitude
of production of $n$ static bosons consists of poles at the energies $E=E_N$
with the residues proportional to $(-1)^N \exp (S_0-s[r])_E$, the latter
being given by the equation (\ref{sef}). The sign alternating factor
$(-1)^N$ arises from $\exp(i I(E_N)/2)= \exp(i \pi N)$ in the numerator of
eq.(\ref{sumt}).  The decay width of the bubble levels can be taken into
account in the Breit-Wigner approximation by shifting the positions of the
poles into the complex plane: $E_N \to E_N - i \Gamma_N/2$.

To conclude this section we collect all the factors in our estimate of the
production amplitudes of $n$ bosons, all being at rest, at energy $E$
corresponding to $n=E/m \gg m^{3-d}/\lambda$ and write the final result in
the form

\beq
\langle n | \phi(0) | 0 \rangle \sim n! \,
{{\exp(i I(E)/2)} \over {1-\exp(i I(E))}} \,
\exp \left [  f_d \,{{d-1} \over d}\, E \, \left ( {E \over {l_d \mu} }
\right )^{1 \over {d-1}} \right ]~.
\label{final}
\eeq
This expression consists of poles corresponding to the energy
levels $E_N$ of quantized bubble, i.e. corresponding to $I(E)= 2 \pi \, N$.
Therefore one can conclude that the production of the bosons in
ultra-high-energy limit goes through intermediate states, which are the
quantum levels of a bubble.

\section{Amplitudes of bubble formation and decay}

The residue of an individual pole in eq.(\ref{final}) is given by

\beq
{\rm res}_N = (-1)^N \, \left ( {E_N \over m} \right )! \, \exp (D(E_N))~,
\label{res}
\eeq
where the shorthand notation $D(E)$ is used for the expression in
eq.(\ref{sef}), which also enters the exponent in eq.(\ref{final}). The
residue can in general be written as

\beq
{\rm res}_N = A(1 \to B_N) \cdot A(B_N \to n)~,
\label{fact}
\eeq
where $A(1 \to B_N)= \langle B_N | \phi(0) | 0 \rangle$ is the amplitude of
production of the $N$-th state of the bubble, $B_N$ by a virtual field
$\phi$ and $A(B_N \to n)$ is the amplitude of transition of this state into
$n$ bosons, all being at rest.  The former amplitude can be evaluated by the
Landau-WKB technique$^{\cite{lwkb}}$ for calculating transition matrix
elements.  According to this technique in the exponential approximation the
matrix element of an operator $f$ between states with energy $E_1=0$ and
$E_2=E$, $\langle E | f | 0 \rangle$, is found by matching the Euclidean
classical trajectories, one with the energy of the initial state, i.e.
$E_1=0$, and the other with the energy of the final state $E_2=E$, which
runs between the matching point and the turning point. The matrix element is
given in the exponential approximation by

\beq
\langle E | f | 0 \rangle \sim \exp (s(E)-s(0))~,
\label{wkbl}
\eeq
where $s(E)$ is the Euclidean-space truncated action on the trajectory.
The specific form of the operator $f$ enters only the pre-exponential
factor (unless the operator $f$
itself is exponential, which is not the case in the problem under
discussion) and can be ignored.

The configuration shown in Fig.3 displays such matching of the evolution in
the Euclidean space. The evolution starts with zero energy, which
corresponds to a flat domain wall. The field then matches that of a
bubble with energy $E$, whose radius then contracts down to the turning
point $r_0$. The configuration is then symmetrically extended beyond the
turning point, so that it gives the square of the matrix element. Clearly,
the difference of the truncated action $(s(E)-s(0))$ on one half of this
configuration, i.e. from $t=-\infty$ to the turning point, is equal to minus
that given by equation (\ref{sef}). Thus one immediately finds the estimate

\beq
|A(1 \to B_N)| =
|\langle B_N |\phi(0) | 0 \rangle | \sim  \exp (-D(E_N))~.
\label{bubam}
\eeq

A comparison of the formulas (\ref{res}), (\ref{fact}) and (\ref{bubam})
leads to the following reasoning. The product (\ref{fact}) of the amplitudes
of formation and decay of the bubble is exponentially large in
$E^{d/(d-1)}$ (eq.(\ref{res})) while the formation amplitude given by
eq.(\ref{bubam}) is exponentially small. Thus the amplitude of the coupling
of the bubble to $n$ bosons, which are all being at rest contains the
doubled positive exponent:

\beq
|A(B_N \to n)| \sim \left ( {E_N \over m} \right )! \, \exp (2 \, D(E_N))~.
\label{bton}
\eeq

To reconcile this extremely strong coupling of the bubble to the state of
$n$ bosons, in which they all have exactly zero spatial momenta, with a
total decay rate that is not exponentially large, one inevitably has to
assume that the coupling of the bubble to bosons develops a form factor
which sharply decreases above the threshold and thus is capable of
suppressing the double-exponential and the factorial energy growth of
the amplitude at the threshold. In view of this observation it is extremely
likely that in the processes $1 \to n$, whose amplitude at the threshold
has only single exponential growth factor, the same form factor makes the
total probability exponentially suppressed at high energy.

\section{Discussion and conclusions}

The results of the search for the correct saddle point for calculation of
the $n$-th coefficient $c_n$ in the expansion of the mean field $\phi$ under
the kink type boundary conditions justify the thin wall approximation used
in this paper in the limits considered. Indeed, the uncertainty of this
approximation, expressed by $\delta$ in eq.(\ref{cq}), which is of the order
of the thickness of the wall, becomes small in comparison with the main term
if the maximal deviation of the wall $h_0$ is much larger than $m^{-1}$. One
can readily see that this is indeed the case in the calculations of this
paper. For the number of space dimensions $d \le 2$ the maximal deviation
grows with the size $R$ of the bounding box and thus for the leading
infrared terms the thin wall approximation is applicable at any $n$. As
mentioned in connection with the equation (\ref{ir}) the result for this
case can be checked against a direct calculation$^{\cite{v2,smith}}$ of the
infrared behavior of the amplitudes at the one-loop level. In the infrared
finite case of $d >2$ in particular for $d=3$ the applicability of the thin
wall approximation is guaranteed in the limit of large $n$: $n \gg
m^{3-d}/\lambda$. This follows from that the maximal deviation $h_0$ is of
the order of $r_0$, thus according to equation (\ref{r0}) $h_0\,m \sim (n\,
\lambda /m^{3-d})^{1/(d-1)}$.

There is possibly not a coincident correlation between the qualitative
dependence of the present calculations on the number of spatial dimensions
and a similar dependence of the spectra of states of many soft bosons with
point-like attraction. Noticing in this connection that the interaction
between soft bosons near the threshold in the theory with the Lagrangian
(\ref{lagr}) is attractive, we recall that for $d \le 2$ an attraction at
short distances produces bound states for any number of bosons, starting
from $n=2$. The existence of such bound states strongly changes the spectral
density of the $n$-particle scattering states, which is another way to
formulate that there is a strong behavior of the form factor for production
amplitudes near the threshold. For $d > 2$, in particular for $d=3$, a weak
attraction at short distances is insufficient to bind few bosons. However
if the number of bosons is non-perturbatively large, $n \gg m^{3-d}/\lambda$
they can form collective multi-particle bound states, which in the
terminology of the present paper are the bubbles of the field $\phi$.
Though a detailed relation to the present calculations is yet to be
understood, this may provide another explanation of why the semiclassical
analysis can be applied for any $n$ if $d \le 2$ and why it becomes
applicable in the case of $d > 2$ only when the number of produced bosons is
non-perturbatively large, $n \gg 1/\lambda$.

The main result of the present paper is the
following hierarchy of the amplitudes for transitions between one
highly virtual particle, the state of $n$ bosons, all having zero spatial
momenta, and a state of a spherical bubble with energy $E=n m$:

$$A(1 \to B) \sim e^{-D(E)} ~,$$
$$A(1 \to n) \sim A(1 \to B) \cdot A(B \to n) \sim e^{D(E)}~, $$
\beq
A(B \to n) \sim e^{2\, D(E)}~.
\label{hier}
\eeq
When the particles have non-zero momenta a form factor arises, due to the
size $r_0$ of the bubble which cuts off the phase space integration. The
suppression due to the form factor can be evaluated from the fact that
the decay rate of a large bubble is not exponentially large in its energy,
i.e. in this exponential scale the rate is $\Gamma (B \to many) \sim O(1)$.
If the processes $1 \to many$ at ultra-high energy are going through the
bubbles, as it is strongly indicated by the calculations in this paper,
their rate should be cut off by the same form factor, which implies

\beq
\Gamma(1 \to many) \sim |A(1 \to B)|^2 \, \Gamma(B \to many) \sim
e^{-2\,D(E)}
\label{estim}
\eeq
and thus these processes are strongly suppressed.

This hierarchy can be extended to processes $many \to many$, which
potentially can go at high temperature. If such scattering processes are
also mediated by the bubbles, one can estimate

\beq
A(n \to n) \sim A(n \to B) \cdot A(B \to n) \sim e^{4\, D(E)}~.
\label{ex4}
\eeq
A kinetical calculation of the rate in a thermal equilibrium in this case
involves the form factor for both the final and the initial states. Thus the
rate of the multi-particle processes at high temperature should be
$\Gamma(many \to many) \sim O(1)$, which is in agreement with
standard thermodynamical calculations.

For a very approximate understanding of the role of the form factor one can
invoke the following reasoning. The amplitude of the process $1 \to n$ above
the threshold is a function of all the $n$ spatial momenta ${\bf p}_i$ of
the final bosons, $A({\bf p}_i)$, and the total probability is given by

\beq
\Gamma = {1 \over {n!}}\int |A({\bf p}_i)|^2 \, \prod_{i=1}^n {{d^d p_i}
\over
{2\varepsilon_i \, (2 \pi)^d}}
\label{totp}
\eeq
with the amplitude $A(0)$ corresponding to all the momenta vanishing being
given by eq.(\ref{final}). The presence of the bubble with the radius $r_0$
in the intermediate state makes the amplitude $A({\bf p}_i)$ to sharply
decrease when any of the momenta ${\bf p}_i$ is of the order of $1/r_0$.
Therefore the phase space corresponding to this region in the momentum of
each particle behaves parametrically as $(r_0)^{-nd} \sim E^{-E\,d/(d-1)}$.
This suppression is sufficient to eliminate the factorial growth of the
amplitude $A(0)$, but is not sufficient to overcome the $\exp (- const \cdot
E^{d/(d-1)})$ enhancement. The estimate of the total rate in
eq.(\ref{estim}) shows that it is most likely that the actual suppression
due to the finite size of the bubble is somewhat stronger than in this
simplistic reasoning, perhaps, due to coherence effects.

As is mentioned before our estimate of the form factor suppression relies on
that the decay rate of a bubble does not grow exponentially as some power of
the energy of the bubble. Therefore it is appropriate to emphasize once
again the arguments in favor of this behavior. One argument is based on the
numerical observation$^{\cite{kobzarev}}$ of the classical relative
stability of the bubble: it dissipates its energy over several oscillations,
thus its lifetime is at least not shorter than of the order of the period of
oscillations $T \sim r_0$. In the quantum theory the suppression of emission
of the bosons from large distances follows from the fact, discussed in
Section 4, that the splitting of the levels of the bubble at high energy is
much smaller than the mass $m$ of the quantum.  Therefore emission of a
boson requires a transition between levels with a large difference $\Delta
N$ of their quantum numbers, which is exponentially suppressed in $\Delta
N$. Thus the emission can take place during the time when the radius of the
bubble is small $r < O(m^{-1})$ and the walls collide into each other.
However even if the probability of annihilation of the walls into outgoing
bosons in this region is of order one, the bubble spends only a small
fraction of time at such small radius, so that the total decay rate is
proportional to the inverse of the period $T$. Therefore there seems to be
every reason to assume that the decay rate of a bubble is limited by
$const/T$.

\section{Acknowledgements}

We are thankful to Brian Smith and Peter Tinyakov for useful
discussions. A.S.G. acknowledges the hospitality of the Theoretical Physics
Institute at the University of Minnesota where this work is done. The work
of M.B.V. is supported in part by the DOE grant DE-AC02-83ER40105.

{\Large \bf Figure captions}\\[0.3 in]
{\bf Figure 1.} A configuration of the field, corresponding to the
inter-phase boundary bent to the maximal deviation $h_0$. The evolution of
the field between the point at a negative time $t$ (heavy dot) and the
boundary proceeds over the time $|t|-h_0 + \delta$, where $\delta$ is of the
order of the thickness of the domain wall. The bold vertical lines at the
edges are the world lines of the boundaries of the spatial bounding
box.\\[0.15 in]
{\bf Figure 2.} The world line of the bubble walls in a
complexified space-time. The bubble evolves in the Euclidean space-time at
$r > r_0$ and in the Minkowski one, when $r < r_0$. \\[0.15 in]
{\bf Figure 3.} The Euclidean field configuration
for calculating the square of the
matrix element $\langle B | \phi | 0 \rangle$ by the Landau-WKB formula.  At
large negative time the field evolves by the classical solution with zero
energy, corresponding to a flat domain wall (lower horisontal line). Then it
matches on the configuration with a large energy $E$, corresponding to a
bubble, which contracts down to the turning radius $r_0$.  Beyond the
turning point the configuration is symmetrically reflected in time, hence it
represents the square of the matrix element. The plus and minus signs
indicate the phases, corresponding to the field approaching $+v$ or $-v$.

\end{document}